\documentclass[aps,preprint]{revtex4}
\usepackage{bm}
\usepackage{amsfonts}
\usepackage{amsmath}
\usepackage{amssymb}
\usepackage{graphicx}%
    \makeatletter
    
    \newcommand{\Rmnum}[1]{\expandafter\@slowromancap\romannumeral #1@}
    \makeatother
\allowdisplaybreaks[4]

\begin{document}
\preprint{CTP-SCU/2012004}
\title{ Vacuum quantum fluctuation energy in expanding universe and dark energy  }
\author{Shun-Jin Wang\\
        \normalsize\textit{Center for Theoretical Physics, College of Physical Science and Technology,}\\
        \normalsize\textit{Sichuan University, Chengdu 610064, China}\\
        \normalsize\textit{E-mail:}
        \texttt{sjwang@home.swjtu.edu.cn}   }

\begin{abstract}
  This article is based on the Planckon densely piled vacuum model and the principle of cosmology. With the Planck era as initial conditions and including the early inflation, we have solved the Einstein-Friedmann equations to describe the evolution of the universe. The results are: 1) the ratio of the dark energy density to the vacuum quantum fluctuation energy density is
  $\frac{{{\rho }_{de}}}{{{\rho }_{vac}}}\sim{{(\frac{{{t}_{P}}}{{{T}_{0}}})}^{2}}\sim{{10}^{-122}} $; 2) at the inflation time ${{t}_{\inf }}={{10}^{-35}}s$, the calculated universe radiation energy density is $\rho ({{t}_{\inf }})\sim{{10}^{-16}}{{\rho }_{vac}}$ and the corresponding temperature is ${{E}_{c}}\sim{{10}^{15}}GeV$ consistent with the GUT phase transition temperature; 3) the expanding universe with vacuum as its environment is a non-equilibrium open system constantly exchanging energy with vacuum; during its expansion, the Planckons in the universe lose quantum fluctuation energy and create the cosmic expansion quanta-cosmons, the energy of cosmons is the lost part of the vacuum quantum fluctuation energy and contributes to the universe energy with the calculated value ${{E}_{\cos mos}}={{10}^{22}}{{M}_{\otimes }}{{c}^{2}}$ (where ${{M}_{\otimes }}$ is solar mass ); 4) the total energy of the universe, namely the negative gravity energy plus the positive universe energy is zero; 5) the negative gravity potential and the gravity acceleration related to the creation of  cosmons are derived with the nature of outward repulsive force, indicating that the cosmon may be the candidate of the dark energy quantum; 6)  both the initial Planck era solution and the infinite asymptotic solution of the Einstein-Friedman equations are unstable: the former tends to expand and the latter tends to shrink, so that the Einstein-Friedman universe will undergo a cyclic evolution of successive expansion and shrinking.
\end{abstract}

\maketitle

\tableofcontents

\section{ Introduction }
The accelerating expansion of the universe and the dark energy are established basic facts in the exact modern cosmology \cite{bib001,bib002}. The existence of vacuum quantum fluctuation energy is also an established basic fact in the modern accurate quantum field theory. Combining the two kinds of basic facts and explaining dark energy as vacuum quantum fluctuation energy is a big step to reach the great goal of unifying quantum theory, relativity, and cosmology \cite{bib003,bib004,bib005,bib006,bib007,bib008}. However, this effort has met serious challenge: the ratio of the dark energy density ${{\rho }_{de}}$ to the vacuum quantum fluctuation energy density ${{\rho }_{vac}}$ shows a huge hierarchy difference of 122 order of magnitude \cite{bib009}: $\frac{{{\rho }_{de}}}{{{\rho }_{vac}}}\sim{{10}^{-122}}$.

  This article is based on the Planckon densely piled vacuum model \cite{bib010} and the principle of cosmology. With the Planck era as initial conditions and including the early inflation, we have solved the Einstein-Friedmann equations to describe the evolution of the universe, a simple and reasonable relation between the dark energy density $\rho_{de}$ and the vacuum quantum fluctuation energy density $\rho_{vac}$ is obtained. The main results are : 1) the solution of Einstein- Friedmann equations has yielded the result
  $\frac{{{\rho }_{de}}}{{{\rho }_{vac}}}\sim{{(\frac{{{t}_{P}}}{{{T}_{0}}})}^{2}}\sim{{10}^{-122}} $ (the Planck time ${{t}_{P}}={{10}^{-43}}s$ and the universe age ${{T}_{0}}={{10}^{18}}s$); 2) at the inflation time ${{t}_{\inf }}={{10}^{-35}}s$, the calculated universe radiation energy density is $\rho ({{t}_{\inf }})\sim{{10}^{-16}}{{\rho }_{vac}}$ and the corresponding temperature is ${{E}_{c}}\sim{{10}^{15}}GeV$ consistent with the GUT phase transition temperature; 3) it is shown that the expanding universe with vacuum as its environment is a non-equilibrium open system constantly exchanging energy with vacuum; during its expanding, the Planckons in the universe lose quantum fluctuation energy and create the cosmic expansion quanta-cosmons, the energy of the cosmons-the lost part of vacuum quantum fluctuation energy contributes to the universe energy with the calculated value ${{E}_{\cos mos}}={{10}^{22}}{{M}_{\otimes }}{{c}^{2}}$ ( where ${{M}_{\otimes }}$ is solar mass ) consistent with the astronomic data; 4) since all Planckons in the vacuum of the expanding universe lose quantum fluctuation energy resulting in negative gravity potential energy and the lost energy of Planckons is used to create cosmons which in turn convert into different kinds of the unverse energy, as the gravity energy is included, the total energy of the universe, namely the negative gravity energy plus the positive universe energy is zero; 5) the negative gravity potential and the gravity acceleration related to the cosmon creation and the simultaneous hole excitation are derived and show the nature of radially outward repulsive force, indicating that the cosmon may be the candidate of the dark energy quantum; 6) it is shown that both the initial boundary solution (the Planck era solution or the Planckon solution) and the infinite asymptotic solution of the Einstein-Friedman equations are unstable: the former tends to expand and the latter tends to shrink, so that the Einstein-Friedman universe undergoes a cyclic evolution of successive expansion and shrinking; 7) solutions to three well known cosmic problems of the Big Bang model is discussed.

     The paper is organized as follows. The basic assumptions and equations are presented in sect.II. Based on the assumptions and equations, the time evolution
  of the universe are studied in detail and analytical expressions for the universe energy density and the universe radius as functions of time, as well as the energy density-radius relation are obtained in sect.III. In sect.IV, based on the obtained solutions, the universe mass and the background microwave temperature are calculated and compared with the astronomic data. In sect.V, based on the Planckon densely piled vacuum model, two kinds of vacuum excitations, namely the radiation excitation due to the Casimir cutoff effect of the supposed universe back hole horizon and the expansion cosmon excitation due to the universe expansion are studied and their gravity effects are compared. An intuitive physical explanation of the obtained results as holographic radiation phenomena is given in sect.VI. Sect.VII is for a summary and discussions of the results, and comparisons of the paper with other theories and models are presented in the final section.

\section{ Basic assumptions and equations }
  The basic assumptions of this article are: (1) the vacuum containing and surrounding the universe as a ubiquitous environment is consisted of densely piled Planckons \cite{bib010}; (2) the universe was born in the Planckon piled vacuum at an explosion of a Planckon in the vacuum, the energy density, radius, and time of the Planckon are thus its  initial conditions(the so-called Planck era condition); (3) the universe obeys the cosmological principle; (4) the evolution of the universe is conducted by the Einstein-Friedmann equations with a flat curvature $k=0$ ( which  can be taken off ) and the energy content includes radiation, cold mater, dark mater, as well as dark energy;  (5) the inflation occurs at the early time ${{\tau }_{\inf }} $.

  The assumptions (1) and (2) are from a study of the microscopic quantum structure of vacuum, black holes, and gravitation systems \cite{bib010}. The assumptions (3) and (4) are based on the commonly accepted physical and cosmological principles \cite{bib003,bib007,bib008}. Since the Einstein-Friedmann equations are of the nature of classical gravitation, the quantum nature of the universe evolution should be incorporated by initial and inflation conditions, namely the assumptions (2) and (5).

  First let us introduce the Planckon to describe the Planck era. As the smallest brick of vacuum, the Planckon is a quantum radiation standing wave in the sphere with the radius ${{r}_{P}}$, volume ${{v}_{P}}$, period ${{\tau }_{P}}$, mass ${{m}_{P}}$, and energy  ${{\varepsilon }_{P}}$ as follows \cite{bib010}:
\begin{equation}\label{equationa003}
       {{r}_{P}}={{(\hbar G/{{c}^{3}})}^{1/2}}\approx c{{\tau }_{P}},~ {{v}_{P}}=\frac{4\pi }{3}{{r}_{P}}^{3},~ {{\tau }_{P}}={{(\hbar G/{{c}^{5}})}^{1/2}}
\end{equation}
\begin{equation}\label{equationa004}
{{m}_{P}}=\frac{1}{2}{{(\hbar c/G)}^{1/2}}=\frac{\hbar }{2c{{r}_{P}}},~ {{\varepsilon }_{P}}={{m}_{P}}{{c}^{2}}=\frac{\hbar c}{2{{r}_{P}}},
\end{equation}
where $ {{\tau }_{P}}\sim{{10}^{-43}}s,\  {{r}_{P}}=c{{\tau }_{P}}\sim{{10}^{-33}}cm,\  {{\varepsilon }_{P}}\sim{{10}^{19}}GeV,\ {{m}_{P}}\sim{{10}^{-5}}g$, and the gravity constant $G=6.67\times {{10}^{-8}}erg\cdot cm/{{g}^{2}}$.

Based on the Planckon densely piled vacuum model \cite{bib010}, the vacuum zero energy density is just the Planckon zero energy density (the Planckon energy ${{\varepsilon }_{P}}$ also corresponds to the cutoff of vacuum quantum fluctuation energy):
\begin{equation}\label{equationa006}
{{\rho }_{vac}}={{\rho }_{P}}={{\varepsilon }_{P}}/{{v}_{P}}={{m}_{P}}{{c}^{2}}/(\frac{4\pi }{3}{{r}_{P}}^{3})=\frac{3{{c}^{4}}}{8\pi G}\frac{1}{{{r}_{P}}^{2}}=\frac{K}{{{(c{{t}_{P}})}^{2}}}
\end{equation}     
\begin{equation}\label{euationa007}
¡¡¡¡¡¡¡¡¡¡¡¡¡¡¡¡¡¡¡¡K=\frac{3{{c}^{4}}}{8\pi G}, {{t}_{P}}=\sqrt{\frac{3{{c}^{2}}}{8\pi G{{\rho }_{vac}}}}. ¡¡
\end{equation}         
According to the above assumptions, the time evolution of the isotropic and homogeneous universe obeys the Einstein-Friedmann equations \cite{bib003}:\\
The expanding equation:
\begin{equation}\label{equationa008}
{{(\frac{{\dot{R}}}{R})}^{2}}=\frac{8\pi G}{3{{c}^{2}}}\rho =\Lambda ={{\lambda }^{2}}, \Lambda =\frac{8\pi G\rho }{3{{c}^{2}}}, \lambda =\sqrt{\frac{8\pi G\rho }{3{{c}^{2}}}}
\end{equation}  
The accelerating equation:
\begin{equation}\label{equationa009}
\frac{{\ddot{R}}}{R}=-\frac{4\pi G}{3{{c}^{2}}}(\rho +3p)
\end{equation}    
  According to the exact moden cosmology, the universe energy density contains the components of radiation ${{\rho }_{r}}$, cold matter ${{\rho }_{m}}$, dark matter ${{\rho }_{dm}}$, and dark energy ${{\rho }_{de}}$. The equation of state is as follows:\\
\begin{equation}\label{equationa010}
     {{p}_{i}}={{w}_{i}}{{\rho }_{i}},~ \rho =\sum\limits_{i}{{{\rho }_{i}}},~ p=\sum\limits_{i}{{{p}_{i}}=\sum\limits_{i}{{{w}_{i}}{{\rho }_{i}}=\sum\limits_{i}{{{w}_{i}}{{x}_{i}}\rho =w\rho }}}
\end{equation}     
\begin{equation}\label{equationa011}
w(t)=\sum\limits_{i}{{{w}_{i}}{{x}_{i}}}(t),~ {{x}_{i}}(t)={{\rho }_{i}}(t)/\rho (t),~ \sum\limits_{i}{{{x}_{i}}(t)=1},~ i=r,m,dm,de
\end{equation}  
\begin{equation}\label{equationa012}
{{w}_{m}}={{w}_{dm}}=0,~ {{w}_{r}}=1/3,~ {{w}_{de}}=-1,~ -1\le w(t)\le 1/3
\end{equation}   

In the following section, we shall obtain the solutions to the Einstein-Friedmann equations, which show that during the evolution of the universe, the energy density components of ${{\rho }_{i}}(t)$, ${{x}_{i}}(t)$, and the coefficient $w_{i}(t)$ of the equations of state vary in their own intervals and make the total energy density $\rho (t)$ vary in time continuously.  Instead, the radial scale factor $R(t)$ of the universe will have the phase change from power law functions to exponential function or the inverse. The Einstein-Friedmann equations can describe both the inflation (exponential) phase and  power law phases  for the radial scale factor function $R(t)$.

\section{ Evolution of the universe }
\subsection{Evolution of the universe energy density $\rho(t)$}
Eq.\eqref{equationa008} and  \eqref{equationa009} lead to
\begin{equation}\label{equationa013}
                      \frac{d\rho }{\sqrt{\rho }(\rho +p)}=-\sqrt{\frac{24\pi G}{{{c}^{2}}}}dt
\end{equation}          
By using the equation of state $p(t)=w(t)\rho(t)$, one has
\begin{equation}\label{equationa014}
          \frac{d\rho }{{{\rho }^{3/2}}}=-\sqrt{\frac{24\pi G}{{{c}^{2}}}}[1+w(t)]dt, ~ d\frac{1}{{{\rho }^{1/2}}}=\sqrt{\frac{6\pi G}{{{c}^{2}}}}[1+w(t)]dt
\end{equation}    
The solution to \eqref{equationa014} is
\begin{equation}\label{equationa015}
             \frac{1}{{{\rho }^{1/2}}(t-{{t}_{0}})}=\frac{1}{{{\rho }^{1/2}}({{t}_{0}})}\{1+\sqrt{\frac{6\pi G\rho ({{t}_{0}})}{{{c}^{2}}}}\int\limits_{{{t}_{0}}}^{t}{[1}+w(\tau )]d\tau \}
\end{equation}   
or
\begin{equation}\label{equationa016}
     \rho (t-{{t}_{0}})=\rho ({{t}_{0}})/\{1+3/2\sqrt{\frac{8\pi G\rho ({{t}_{0}})}{3{{c}^{2}}}}{{[(t-{{t}_{0}})+\int\limits_{{{t}_{0}}}^{t}{w(\tau )d\tau ]}\}^2}}
\end{equation}
In viewing the interval $[-1,\frac{1}{3}]$ of $w(t)$ and the $(t-{{t}_{0}})$-term in eq. \eqref{equationa016},  $\rho (t-{{t}_{0}})$ in general, is not so sensitive to $w(t)$ (only when $w(t)\to -1$, $\rho (t-{{t}_{0}})$ changes from inverse quadrature function to constant). In contrast, $R(t-{{t}_{0}})$ is very sensitive to $w(t)$ and can change from power functions to exponential function. This behaviour leads to the problem of consistence between $\rho (t)$ and $R(t)$ as any approximation is made. The evolution detail of different energy density components ${{\rho }_{i}}(t)$ controlling the evolution of $w(t)$, is very important to produce a physically reasonable and realistic evolution for both $\rho (t)$ and $R(t)$, and to establish the consistence between $\rho (t)$ and $R(t)$. Since the evolution of ${{\rho }_{i}}(t)$ is related to the interactions of elementary particles  under astronomic and cosmic conditions, the knowledge of particle physics is needed for a detailed description.

If the initial conditions are the Planck era, namely  ${{t}_{0}}={{t}_{P}}=\sqrt{\frac{3{{c}^{2}}}{8\pi G{{\rho }_{vac}}}}$ and $\rho ({{t}_{P}})={{\rho }_{vac}}$, the solution is
\begin{eqnarray}
 \rho (t-{{t}_{P}})&=&{{\rho }_{vac}}/\{1+(3/2)\frac{1}{{{t}_{P}}}{{[(t-{{t}_{P}})+\int\limits_{{{t}_{P}}}^{t}{w(\tau )d\tau ]}\}^2}} \nonumber\\
 &=&{{\rho }_{vac}}/{{[1+(3/2)\frac{(t-{{t}_{P}})}{{{t}_{P}}}(1+{{w}_{mid}}(t))]^2}}
\end{eqnarray}

${{w}_{mid}}(t)$ is the integration mid value of $w(t)$ from ${{t}_{P}}$ to $t$: $\int\limits_{{{t}_{P}}}^{t}{w(\tau )d\tau ={{w}_{mid}}(t)(t-{{t}_{P}})}$.¡¡\\¡¡
     For the dark energy:
     \begin{gather}
     {{w}_{de}}(t)=-1,~ {{\rho }_{de}}(t-{{t}_{P}})={{\rho }_{vac}} ;
      \end{gather}        
     For the radiation:
     \begin{gather}
      {{w}_{r}}(t)=\frac{1}{3},~¡¡{{\rho }_{r}}(t-{{t}_{P}})={{\rho }_{vac}}/{{[1+2(t-t{}_{P})/{{t}_{P}}]}^{2}} ;
      \end{gather}   
     For the cold matter:
     \begin{gather}
    {{w}_{m}}(t)=0,~ {{\rho }_{m}}(t-{{t}_{P}})={{\rho }_{vac}}/{{[1+\frac{3}{2}(t-t{}_{P})/{{t}_{P}}]}^{2}} ;
     \end{gather}  
    For the global average evolution from $t_P$ to present time $T_0$ (for its definition, see below):
    \begin{gather}
   {{w}_{mid}}(t)=-1/3,~ {\rho }(t-{{t}_{P}})={{\rho }_{vac}}/{{[1+(t-t{}_{P})/{{t}_{P}}]}^{2}} .
    \end{gather} 

   ¡¡According to the modern cosmology, the creation and evolution of the universe contain three stages.  Let us consider the detail of the evolution in the corresponding three steps so that the inflation can be incorporated in the early time. \\
1) From Planck era (${{t}_{P}},{{r}_{P}},{{\rho }_{P}}={{\rho }_{vac}}$) to inflation starting moment[${{\tau }_{\inf }},R({{\tau }_{\inf }}),\rho ({{\tau }_{\inf }})$],
     the evolution is of radiation type: $w(t)=1/3$; \\
2) From[${{\tau }_{\inf }},~R({{\tau }_{\inf }}),~\rho ({{\tau }_{\inf }})$] to [${{t}_{\inf }},~R({{t}_{\inf }}),~\rho ({{t}_{\inf }})=\rho ({{\tau }_{\inf }})$], the evolution is of inflation: $w(t)=-1$;\\
3) From [${{t}_{\inf }},~R({{t}_{\inf }}),~\rho ({{t}_{\inf }})$] to the present universe [${{T}_{0}},~{{R}_{0}},~\rho ({{T}_{0}})$], the evolution is of a type of the  mixture of radiation, matter, and dark energy.

  The specific evolutions of three steps are in the following:

(1) From ${{t}_{P}}={{10}^{-43}}s$ to inflation starting moment $t={{\tau }_{\inf }}\sim{{10}^{-35}}s$(${{w}_{mid}}=1/3$).
The solution is
\begin{eqnarray}
   \rho ({{\tau }_{\inf }}-{{t}_{P}})&=&{{\rho }_{vac}}/\{1+\frac{3}{2{{t}_{P}}}{{[({{\tau }_{\inf }}-{{t}_{P}})+\int\limits_{{{t}_{P}}}^{{{\tau }_{\inf }}}{w(\tau )d\tau ]}\}}^{2}}\approx \frac{4}{9}{{\rho }_{vac}}{{[\frac{{{t}_{P}}}{({{\tau }_{\inf }}-{{t}_{P}})(1+{{w}_{mid}})}]}^{2}} \nonumber \\
 &=&\frac{1}{4}{{\rho }_{vac}}{{(\frac{{{t}_{P}}}{{{\tau }_{\inf }}-{{t}_{P}}})}^{2}}
\end{eqnarray}     

The energy density before inflation at ${{\tau }_{\inf }}$ is
\begin{gather}
\frac{\rho ({{\tau }_{\inf }}-{{t}_{P}})}{{{\rho }_{vac}}}=\frac{1}{4}{{[\frac{{{t}_{P}}}{({{\tau }_{\inf }}-{{t}_{P}})}]}^{2}}\approx {{10}^{-16}}
\end{gather}             
The radiation energy density $\rho ({{\tau }_{\inf }}-{{t}_{P}})$, the radiation quantum energy ${{E}_{c}}$ and its wave length ${{\lambda}_{c}}$ before inflation, and the Planckon energy ${{\varepsilon}_{P}}$ and its wave length ${{\lambda}_{P}}$ have the relation:
\begin{equation}
  \rho ({{\tau }_{\inf }})/{{\rho }_{vac}}={{(\frac{{{\lambda }_{P}}}{{{\lambda }_{c}}})}^{4}}={{(\frac{{{r}_{P}}}{r{}_{c}})}^{4}}={{(\frac{{{E}_{c}}}{{{\varepsilon}_{P}}})}^{4}}
  \end{equation}
where ${{\lambda }_{c}}\sim{{r}_{c}}$ and ${{\lambda }_{P}}\sim{{r}_{P}}$  are the corresponding  wave  lengths.

From ${{\varepsilon}_{P}}={{10}^{19}}GeV$, one obtains
\begin{gather}
 \rho ({{\tau }_{\inf }})/{{\rho }_{vac}}={{(\frac{{{E}_{c}}}{{{\varepsilon}_{P}}})}^{4}}={{(\frac{{{E}_{c}}}{{{10}^{19}}GeV})}^{4}}\sim{{10}^{-16}},{{E}_{c}}\sim{{10}^{15}}GeV
\end{gather} 
which is the temperature of the GUT at the time of phase transition and indicates that the first step evolution from ${{t}_{P}}$ to ${{\tau }_{\inf }}$ is reasonable and yields correct result.

(2) From [${{\tau }_{\inf }},R({{\tau }_{\inf }}),\rho ({{\tau }_{\inf }})$] to [${{t}_{\inf }}={{10}^{-34}}s,R({{t}_{\inf }}),\rho ({{t}_{\inf }})=\rho ({{\tau }_{\inf }})$]: the dark energy( ${{w}_{mid}}=-1$ ) controlling evolution  leads to  $\rho ({{t}_{\inf }})=\rho ({{\tau }_{\inf }})$.

(3) From [${{t}_{\inf }}={{10}^{-34}}s$, $\rho ({{t}_{\inf }})$] to the present time ${{T}_{0}}={{10}^{18}}s$, the solution is
\begin{eqnarray}
\rho ({{T}_{0}}-{{t}_{\inf }})&=&\rho ({{t}_{\inf }})/\{1+(3/2)\frac{1}{{{{\bar{t}}}_{\inf }}}{{[({{T}_{0}}-{{t}_{\inf }})+\int\limits_{{{t}_{\inf }}}^{{{T}_{0}}}{w(\tau )d\tau ]}\}^2}}\approx \frac{4}{9}{{\rho }_{\inf }}\frac{{{{\bar{t}}}_{\inf }}^{2}}{[({{T}_{0}}-{{t}_{\inf }})+\int\limits_{{{t}_{\inf }}}^{{{T}_{0}}}{w(\tau )d\tau {{]}^{2}}}} \nonumber\\
  &=&\frac{4}{9}\rho ({{t}_{\inf }}){{[\frac{{{{\bar{t}}}_{\inf }}}{({{T}_{0}}-{{t}_{\inf }})(1+{{w}_{mid}})}]^2}} \nonumber \\
 \frac{\rho ({{T}_{0}}-{{t}_{\inf }})}{\rho ({{t}_{\inf }})}&=&\frac{4}{9}{{[\frac{{{{\bar{t}}}_{\inf }}}{({{T}_{0}}-{{t}_{\inf }})(1+{{w}_{mid}})}]^2}}\approx {{10}^{-106}}, ~~~{{\bar{t}}_{\inf }}=\sqrt{\frac{3{{c}^{2}}}{8\pi G\rho ({{t}_{\inf }})}}\approx {{10}^{-35}}s
 \end{eqnarray}  
 where ${{w}_{mid}} $ is the mid value of $w(t)$ from ${{t}_{\inf }}={{10}^{-34}}s$ to ${{T}_{0}}={{10}^{18}}s$.  In the last subsection, we will know that
the evolution of $R(t)$ requires that ${{w}_{mid}}=-1/3$.  From $\rho ({{\tau }_{\inf }})=\rho ({{t}_{\inf }})$ and $\rho ({{\tau }_{\inf }})/{{\rho }_{vac}}\sim{{10}^{-16}}$, one gets
 \begin{gather}
    \rho ({{T}_{0}}-{{t}_{\inf }})=\rho ({{t}_{\inf }}){{10}^{-106}}={{\rho }_{vac}}\frac{\rho ({{t}_{\inf }})}{{{\rho }_{vac}}}{{10}^{-106}}={{\rho }_{vac}}\frac{\rho ({{\tau }_{\inf }})}{{{\rho }_{vac}}}{{10}^{-106}}={{\rho }_{vac}}{{10}^{-122}}
 \end{gather}      
and $\frac{\rho ({{T}_{0}}-{{t}_{\inf}})}{{{\rho }_{vac}}}\approx {{10}^{-122}}$. As ${{t}_{\inf }}={{10}^{-33}}s$, the result is the same since the value of
${{t}_{\inf }}<<{{T}_{0}}$ yields negligible effect in (${{T}_{0}}-{{t}_{\inf }}$).

Now consider one step evolution from ${{t}_{P}}$ directly to ${{T}_{0}}={{10}^{18}}s$: the solution is
\begin{eqnarray}\label{equationa028}
\rho ({{T}_{0}}-{{t}_{P}})&=&{{\rho }_{vac}}/\{1+(3/2)\frac{1}{{{t}_{P}}}{{[({{T}_{0}}-{{t}_{P}})+\int\limits_{{{t}_{P}}}^{{{T}_{0}}}{w(\tau )d\tau ]}\}^2}} \nonumber \\
&\approx& \frac{4}{9}{{\rho }_{vac}}\frac{{{t}_{P}}^{2}}{[({{T}_{0}}-{{t}_{P}})+\int\limits_{{{t}_{P}}}^{{{T}_{0}}}{w(\tau )d\tau {{]}^{2}}}}=\frac{4}{9}{{\rho }_{vac}}{{[\frac{{{t}_{P}}}{({{T}_{0}}-{{t}_{P}})(1+{{w}_{mid}})}]^2}}
\end{eqnarray}

The ratio of the universe energy density to the vacuum energy density is
\begin{gather}\label{equationa029}
\frac{\rho ({{T}_{0}}-{{t}_{P}})}{{{\rho }_{vac}}}=\frac{4}{9}{{[\frac{{{t}_{P}}}{({{T}_{0}}-{{t}_{P}})(1+{{w}_{mid}})}]}^{2}}\approx {{10}^{-122}}
\end{gather}£®             
where ${{w}_{mid}}$ is the mid value between ${{t}_{P}}$ and ${{T}_{0}}$. Since one step evolution(${{t}_{P}}\to{{T}_{0}}$) and three step evolution(${{t}_{P}}\to {{\tau }_{\inf }}\to {{t}_{\inf }}\to {{T}_{0}}$) yield the same result, the time duration of phase transition thus has no effect on the energy density of the universe at the  present time.  It should be pointed out the initial condition of the Planck era is a key point to obtain the correct results, indicating that \textbf{ our universe was created and started at an explosion of a Planckon in the Planckon piled vacuum, not from a geometric point.}

Since at the present time ${{T}_{0}}$,  the dark energy density ${{\rho }_{de}}$  is  $\textit{73}\% $ of  the universe total energy density, so we have
\begin{gather}
          \rho ({{T}_{0}})\approx {{\rho }_{de}}({{T}_{0}}) ,~~~~   \frac{{{\rho }_{de}}}{{{\rho }_{vac}}}={{10}^{-122}}
\end{gather}                     

 To study the consistence between $\rho (t)$ and $R(t)$ and other cosmic problems, we should investigate the radial scale factor $R(t)$ evolution.
\subsection{ Evolution of the universe radius $R(t)$ }
 From the expansion equation, we have the equation for $R(t)$:
 \begin{equation}
 \frac{dR}{R}=\sqrt{\frac{8\pi G\rho (t)}{3{{c}^{2}}}}dt=\lambda (t)dt, ~~ \sqrt{\frac{8\pi G\rho (t)}{3{{c}^{2}}}}=\lambda(t)
\end{equation}

The universe energy density $\rho(t)$ has been obtained before as follows
\begin{equation}
\rho (t-{{t}_{0}})=\rho (t_0)/\{1+(3/2)\sqrt{\frac{8\pi G\rho ({{t}_{0}})}{3{{c}^{2}}}}[(t-{{t}_{0}})+\int\limits_{t_0}^{t}\omega(\tau )d\tau ]\}^2
\end{equation}

The exact solution of $R(t)$ is
\begin{equation}
R(t)=R({{t}_{0}})\exp [\int\limits_{{{t}_{0}}}^{t}{\lambda (\tau )d\tau ]}
\end{equation}                     

The integration in the exponential can be calculated by substitution of $\rho (t)$,
\begin{equation}
\int\limits_{{{t}_{0}}}^{t}{\lambda (\tau )d\tau =\sqrt{\frac{8\pi G\rho ({{t}_{0}})}{3{{c}^{2}}}}}\int\limits_{{{t}_{0}}}^{t}{\frac{d\tau }{1+3/2\sqrt{\frac{8\pi G\rho ({{t}_{0}})}{3{{c}^{2}}}}[(\tau -{{t}_{0}})+\int\limits_{{{t}_{0}}}^{\tau }{w(\tau ')d\tau ']}}}
\end{equation}    

For the dark energy dominant $w(t)=-1$, it follows
\begin{equation}
\int\limits_{{{t}_{0}}}^{t}{\lambda (\tau )d\tau =\sqrt{\frac{8\pi G\rho ({{t}_{0}})}{3{{c}^{2}}}}}(t-{{t}_{0}})
\end{equation}    
\begin{equation}
R(t)=R({{t}_{0}})\exp [\sqrt{\frac{8\pi G\rho ({{t}_{0}})}{3{{c}^{2}}}}(t-{{t}_{0}})]
\end{equation}           
which is of radial inflation.

With mid value theorem $\int\limits_{{{t}_{0}}}^{t}{w(\tau )d\tau ={{w}_{mid}}(t)(t-{{t}_{0}})}$, one has
\begin{gather}
\int\limits_{{{t}_{0}}}^{t}{\lambda (\tau )d\tau =}\ln {{\{1+\frac{3(1+{{w}_{mid}(t)})}{2}[\alpha (t)-\alpha ({{t}_{0}})]\}}^{\frac{2}{3(1+{{w}_{mid}(t)})}}},~~
\alpha (t)=\sqrt{\frac{8\pi G\rho ({{t}_{0}})}{3{{c}^{2}}}}t
\end{gather}  
The solution in terms of ${{w}_{mid}(t)}$ is
\begin{equation}
R(t)=R({{t}_{0}}){{[1+\frac{1}{\frac{2}{3(1+{{w}_{mid}(t)})}}\sqrt{\frac{8\pi G\rho ({{t}_{0}})}{3{{c}^{2}}}}(t-{{t}_{0}})]}^{\frac{2}{3(1+{{w}_{mid}(t)})}}}
\end{equation}   
which is very sensitive to  ${{w}_{mid}(t)}$.  As $t$ is fixed at some value $T$, ${{w}_{mid}(T)}$ becomes the mid value ${{w}_{mid}}$ in the time interval
$[t_{0},T]$ as discussed before.

1)  As ${{w}_{mid}}\to -1$ (dark energy dominant) $\frac{2}{3(1+{{w}_{mid}})}\to \infty $,
\begin{eqnarray}\label{equationa039}
&R(t)& = R({{t}_{0}}){{[1+\frac{1}{\frac{2}{3(1+{{w}_{mid}})}}\sqrt{\frac{8\pi G\rho ({{t}_{0}})}{3{{c}^{2}}}}(t-{{t}_{0}})]}^{\frac{2}{3(1+{{w}_{mid}})}}}\to \nonumber \\
&R(t)& =R({{t}_{0}})\exp [\sqrt{\frac{8\pi G\rho ({{t}_{0}})}{3{{c}^{2}}}}(t-{{t}_{0}})]
\end{eqnarray}  

Eq.\eqref{equationa039} is the same as that obtained before.

The velocity after inflation: from ${{\tau }_{\inf }}=\sqrt{\frac{3{{c}^{2}}}{8\pi G\rho ({{\tau }_{\inf }})}}={{10}^{-35}}s$  and  ${{t}_{\inf }}={{10}^{-33}}s$,   one  gets
\begin{eqnarray}
 &\dot{R}&({{t}_{\inf }}={{10}^{-33}}s)=(R({{\tau }_{\inf }}={{10}^{-35}}s)/{{\tau }_{\inf }})\times {{e}^{({{t}_{\inf }}-{{\tau }_{\inf }})/{{\tau }_{\inf }}}}= R({{\tau }_{\inf }}={{10}^{-35}}s)/{{\tau }_{\inf }}\times {{10}^{43}}  \nonumber \\
 &=&({{10}^{-29}}/{{10}^{-35}}){{10}^{43}}cm/s\sim{{10}^{49}}cm/s \quad ({{e}^{100}}\approx {{10}^{43}}),
\end{eqnarray}
 the effect of curvature $k$ term approaches  zero:  $ \frac{\rho -{{\rho }_{c}}}{{{\rho }_{c}}}=\frac{{{c}^{2}}k}{{{H}^{2}}{{R}^{2}}}=\frac{{{c}^{2}}k}{{{{\dot{R}}}^{2}}}\sim{{10}^{-78}}\sim0$, so  after inflation the space becomes flat. We will show later, before inflation, the effect of curvature $k$ term is very large and the space is not flat.

 2)  As  ${{w}_{mid}}\to 1/3$ (radiation dominant)
\begin{equation}
R(t)=R({{t}_{0}}){{[1+2\sqrt{\frac{8\pi G\rho ({{t}_{0}})}{3{{c}^{2}}}}(t-{{t}_{0}})]^{1/2}}},
\end{equation} 
The radial velocity at ${{t}_{P}}$ and ${{\tau }_{\inf }}$:
\begin{equation}
\dot{R}(t)=\frac{{{r}_{P}}}{{{t}_{P}}}{{[1+2(t-{{t}_{P}})/{{t}_{P}}]}^{-1/2}}\approx {{10}^{10}}\times {{[1+2(t-{{t}_{P}})/{{t}_{P}}]}^{-1/2}}cm/s,
\end{equation}
The velocity at $t\approx {{t}_{P}}$: $\dot{R}(t\sim{{t}_{P}})\sim{{10}^{10}}cm/s\sim c$, \\
The velocity before inflation at $t={{\tau }_{\inf }}={{10}^{-35}}s$: $\dot{R}({{\tau }_{\inf }})\sim{{10}^{6}}cm/s$.\\
The curvature effect before inflation is very lage: $\frac{\rho -{{\rho }_{c}}}{{{\rho }_{c}}}=\frac{{{c}^{2}}k}{{{H}^{2}}{{R}^{2}}}=\frac{{{c}^{2}}k}{{{{\dot{R}}}^{2}}}\approx {{10}^{8}}>>1$, \\
so the space is not flat.

3)  As ${{w}_{mid}}\to 0 $(cold matter dominant)
\begin{equation}
R(t)=R({{t}_{0}}){{[1+\frac{3}{2}\sqrt{\frac{8\pi G\rho ({{t}_{0}})}{3{{c}^{2}}}}(t-{{t}_{0}})]^{\frac{2}{3}}}}
\end{equation}  

4) The global average evolution from ${{t}_{P}}$ to ${{T}_{0}}$: in the last subsection, we shall show that Nature requires ${{w}_{mid}}=-\frac{1}{3}$ and
the global average evolution of $R(t)$ is
\begin{equation}
R(t)=R({{t}_{0}}={{t}_{P}})[1+\sqrt{\frac{8\pi G\rho ({{t}_{0}})}{3{{c}^{2}}}}(t-{{t}_{0}})]={{r}_{P}}[1+(t-{{t}_{P}})/{{t}_{P}}],
\end{equation} 
the evolution is linear.

The radial scales of the universe  at ${{\tau }_{\inf }}={{10}^{-35}}s$ and ${{t}_{\inf }}={{10}^{-33}}s$:\\
(1) The radiation dominant(${{w}_{mid}}=1/3$),  from (${{t}_{P}},{{r}_{P}},{{\rho }_{vac}}$) to
      (${{\tau }_{\inf }}={{10}^{-35}}s,R({{\tau }_{\inf }}),\rho ({{\tau }_{\inf }})$):
\begin{equation}
¡¡¡¡R({{\tau }_{\inf }})=R({{t}_{P}}){{[1+2\sqrt{\frac{8\pi G\rho ({{t}_{P}})}{3{{c}^{2}}}}(t-{{t}_{P}})]}^{1/2}}=\sqrt{2}{{r}_{P}}{{(\frac{{{\tau }_{\inf }}-{{t}_{P}}}{{{t}_{P}}})}^{1/2}}\approx {{10}^{-29}}cm.
\end{equation}
(2)  Inflation(${{w}_{mid}}=-1$), from(${{\tau }_{\inf }}, R({{\tau }_{\inf }}), \rho ({{\tau }_{\inf }})$) to (${{t}_{\inf }}={{10}^{-33}}s, R({{t}_{\inf }})$,  $\rho ({{t}_{\inf }})$):\\ from
     ${{t}_{0}}={{\tau }_{\inf }}$, $t={{t}_{\inf }}={{10}^{-33}}s$, $\rho ({{t}_{0}})=\rho ({{\tau }_{\inf }})$, $\sqrt{8\pi \rho ({{\tau }_{\inf }})/3{{c}^{2}}}=1/{{\tau }_{\inf }}$,
              ${{\tau }_{\inf }}={{10}^{-35}}s$, and $R({{\tau }_{\inf }})={{10}^{-29}}cm$, one obtains
\begin{eqnarray}
  R({{t}_{\inf }})&=&R({{\tau }_{\inf }})\exp [\sqrt{\frac{8\pi G\rho ({{\tau }_{\inf }})}{3{{c}^{2}}}}({{t}_{\inf }}-{{\tau }_{\inf }})]={{10}^{-29}}\times {{e}^{\frac{{{t}_{\inf }}}{{{\tau }_{\inf }}}}}cm
={{10}^{-29}}\times {{10}^{\lg e\cdot 100}}cm \nonumber \\
   &=&{{10}^{-29}}\times {{10}^{43}}cm={{10}^{14}}cm \nonumber \\
   (\text{if } {{t}_{\inf }}&=&{{10}^{-33.161}}s,{{\tau }_{\inf }}={{10}^{-35}}s,\text{then } R({{t}_{\inf }})=10cm).
\end{eqnarray}

\subsection{ The $\rho(R)-R$ relation }
The evolution equation for $R(t)$ is related to energy conservation as follows :
\begin{equation}\label{equationa048}
               \frac{d(\rho {{R}^{3}})}{dt}=-p\frac{d({{R}^{3}})}{dt}.
\end{equation}                        
Physical implication of eq.\eqref{equationa048}: as $p>0$ (radiation dominant),  $\frac{d({{R}^{3}})}{dt}>0$, and $\frac{d(\rho {{R}^{3}})}{dt}<0$, the expanding universe loses energy to vacuum; as $p<0$ (dark energy dominant), $\frac{d({{R}^{3}})}{dt}>0$, and $\frac{d(\rho {{R}^{3}})}{dt}>0$, the expanding universe acquires energy from vacuum. Furthermore, from the above equations and discussions, one can find that as $R$ approaches infinite, the solution of the Einstein-Friedman equations is unstable with respect to the perturbation of decreasing $R$ and the universe will start to shrink inwards due to positive feedback effect. On the other hand, for the initial state solution, namely the Planck era solution or the Planckon solution, the compression of the Planckon will make its energy density and pressure larger than those of its environment consisting of Planckons,  a rebound will be resulted and the afterwards expansion of the universe will start and be conducted by the Einstein-Friedman equations. In other words, the initial state solution is also unstable and tends to expand. Therefore, the  Einstein-Friedman universe with the Planckon solution as its initial condition in its expansion phase and as its smallest limit in its shrinking phase, will undergo a cyclic evolution of successive expansion and shrinking.

The total energy $E$ of the universe: From $p=w\rho $, $E=\rho V$, and $V=\frac{4\pi }{3}{{R}^{3}}$, one obtains
\begin{equation}
d\ln [E]=-wd\ln V
\end{equation}
If $w$ assumes the constant mid value $w={{w}_{mid}}=const$, then
\begin{equation}
d\ln [E{{V}^{{{w}_{mid}}}}]=0,¡¡E=C/{{V}^{{{w}_{mid}}}}
\end{equation}
The total energy evolution with $R$ are as follows:

Radiation dominant(${{w}_{mid}}=1/3$):
\begin{equation}
{{E}_{r}}=C/R
\end{equation}

Cold matter dominant(${{w}_{mid}}=0$ ):

\begin{equation}
{{E}_{m}}=C
\end{equation}

Dark energy dominant(${{w}_{mid}}=-1$):
\begin{equation}
{{E}_{de}}=C{{R}^{3}}
\end{equation}

Global average evolution(${{w}_{mid}}=-1/3$):
\begin{equation}
\bar{E}=CR
\end{equation}

  The above evolutions of energy density and total energy of the universe indicate that the expanding universe with the Planckon piled vacuum as its environment is a non-equilibrium open system constantly exchanging energy with the vacuum. The universe was born in the Planckon piled vacuum at an explosion of a Planckon of the vacuum, the Planck era, namely the radius, time, and energy density of the Planckon constitutes its initial condition. During its expansion, more and more Planckons are involved in the universe evolution and all the involved Planckons in the universe lose their quantum fluctuation energy and the lost irregular quantum fluctuation energy of Planckons converts into cosmons and transforms into different kinds of universe energy. The expanding universe acquires energy from vacuum in the form of cosmons which are created and made of the lost quantum fluctuation energy of Planckons. As shown later, the cosmons are  the candidate of the dark energy quanta with the nature of repulsive gravity force. The cold matter has no net energy exchange with vacuum on average.

  Combining the energy conservation equation and the radial evolution equation,  one has
\begin{equation}
¡¡\frac{d\rho }{{{\rho }^{3/2}}}=-\sqrt{\frac{24\pi G}{{{c}^{2}}}}[1+w(t)]dt,~~~\frac{dR}{R}=\sqrt{\frac{8\pi G\rho (t)}{3{{c}^{2}}}}dt=\lambda (t)dt
\end{equation}

and
\begin{equation}
\frac{dR}{R}=-\frac{d\rho }{3[1+w(\rho )]\rho },~~~\ln \frac{R}{{{r}_{P}}}=\int\limits_{\rho }^{{{\rho }_{P}}}{\frac{d\rho }{3[1+w(\rho )]\rho }=\frac{1}{3[1+w({{\rho }_{mid}})]}}\ln \frac{{{\rho }_{P}}}{\rho }
\end{equation}
¡¡¡¡
Their solutions yield the $\rho =\rho (R)$-$R$ relation:
\begin{equation}
\rho (R)={{\rho }_{vac}}{{(\frac{{{r}_{P}}}{R})}^{3(1+{{w}_{mid}})}} ~~~({{\rho }_{P}}={{\rho }_{vac}},~ w_{mid}=w(\rho_{mid}) )
\end{equation}
¡¡¡¡¡¡¡¡¡¡¡¡¡¡
Different ${{w}_{mid}}$ lead  to different power laws:

Radiation dominant:
\begin{equation}
{{w}_{mid}}=1/3,~ \rho ={{\rho }_{vac}}{{(\frac{{{r}_{P}}}{R})}^{4}}
\end{equation}

Matter dominant:
\begin{equation}
{{w}_{mid}}=0,~¡¡\rho ={{\rho }_{vac}}{{(\frac{{{r}_{P}}}{R})}^{3}},
\end{equation}

Dark energy dominant:
\begin{equation}
{{w}_{mid}}=-1,~¡¡\rho ={{\rho }_{vac}}
\end{equation}

Global average evolution:¡¡
\begin{equation}
{{w}_{mid}}=-1/3,~ \rho ={{\rho }_{vac}}{{(\frac{{{r}_{P}}}{R})}^{2}}
\end{equation}

Summary of the evolution phase transitions:\\
  $\rho (t)-t$~ function: sensitive to $w(t)$,   transitions within power laws; \\
  $R(t)-t$~ function : very sensitive to $w(t)$, transitions between power laws and exponential; \\
¡¡$\rho (R)-R$~ function: sensitive to $w(t)$,  transitions within different power laws.

\subsection{More information from $\rho (t)$ and $R(t)$}
From the evolution of $\rho (t)$ and $R(t)$,  one can obtain more information.~The value ${{w}_{mid}}>-1$ definitely leads to $\frac{\rho ({{T}_{0}})}{{{\rho }_{vac}}}\sim{\ }{{10}^{-122}}$ (see eq. \eqref{equationa028} and eq.\eqref{equationa029}). From $R(t-{{t}_{P}})$ versus $ (t-{{t}_{P}})$  relation:
\begin{equation}
          R(t)=R({{t}_{P}}){{[1+\frac{1}{\frac{2}{3(1+{{w}_{mid}})}}\sqrt{\frac{8\pi G\rho ({{t}_{P}})}{3{{c}^{2}}}}(t-{{t}_{P}})]}^{\frac{2}{3(1+{{w}_{mid}})}}}
\end{equation}
\begin{equation}
R({{t}_{P}})={{r}_{P}},~ \rho ({{t}_{P}})={{\rho }_{vac}},~ \sqrt{8\pi G{{\rho }_{vac}}/3{{c}^{2}}}=1/{{t}_{P}}
\end{equation}
one can derive that if $R({{T}_{0}})={{10}^{28}}cm$ and $\ddot{R}>0$  today,  then there must be
 ${{w}_{mid}}\approx -1/3$ for the global average evolution, namely:
  $R({{T}_{0}})={{10}^{28}}cm\to {{w}_{mid}}\sim-\frac{1}{3}$ and  $\ddot{R}>0\to {{w}_{mid}}<-\frac{1}{3}$.

 In fact, as ${{w}_{mid}}\approx -1/3$, $\frac{2}{3(1+{{w}_{mid}})}\approx 1$, $R(T{}_{0}-{{t}_{P}})={{r}_{P}}[1+({{T}_{0}}-{{t}_{P}})/{{t}_{P}}]\approx {{10}^{-33}}\times {{10}^{61}}cm={{10}^{28}}cm$  with ${{r}_{P}}={{10}^{-33}}cm$, ${{t}_{P}}={{10}^{-43}}s$, and ${{T}_{0}}={{10}^{18}}s$.
   As ${{w}_{mid}}\approx -\frac{1}{3}$ and $\frac{2}{3(1+{{w}_{mid}})}\approx 1 $, then $R(T{}_{0}-{{t}_{P}})={{r}_{P}}[1+({{T}_{0}}-{{t}_{P}})/{{t}_{P}}]$ which yields: the acceleration $\ddot{R}({{T}_{0}})\sim{{10}^{-7}}cm/{{s}^{2}}$ and velocity $\dot{R}({{T}_{0}})\sim{{10}^{10}}cm/s\sim c $ at the present time.

The following calculations of $R({{T}_{0}})$ indicate  that any single component energy density can not yield the observed value of $R({{T}_{0}})$:

Radiation dominant: ${{w}_{mid}}\to 1/3 $,
$R(t)=R({{t}_{0}})[1+2\sqrt{\frac{8\pi G\rho (t_0)}{3{{c}^{2}}}}(t-{{t}_{0}})]^{1/2},~ R(T_0)\sim 10^{-3}cm \ll 10^{28}cm=R_0,$
 the value is too small;

Matter dominant: ${{w}_{mid}}\to 0$,
$R(t)=R({{t}_{0}}){{[1+\frac{3}{2}\sqrt{\frac{8\pi G\rho ({{t}_{0}})}{3{{c}^{2}}}}(t-{{t}_{0}})]}^{{2}/{3}}}$,
                           $R({{T}_{0}})\sim{{10}^{7}}cm\ll{{10}^{28}}cm={{R}_{0}}$,
the value is also too small;

Dark energy dominant: ${{w}_{mid}}\to -1$,
\begin{eqnarray*}
&R(t)&=R({{t}_{0}}){{[1+\frac{1}{\frac{2}{3(1+{{w}_{mid}})}}\sqrt{\frac{8\pi G\rho ({{t}_{0}})}{3{{c}^{2}}}}(t-{{t}_{0}})]}^{\frac{2}{3(1+{{w}_{mid}})}}}\to \\
&R(t)&=R({{t}_{0}})\exp [\sqrt{\frac{8\pi G\rho ({{t}_{0}})}{3{{c}^{2}}}}(t-{{t}_{0}})]  \\
&R({{T}_{0}})&={{r}_{P}}{{e}^{{{T}_{0}}/{{t}_{P}}}}\approx {{10}^{-33 + 0.43 \times {{10}^{61}}}}cm\gg{{10}^{28}}cm={{R}_{0}},
\end{eqnarray*}
the value is too large.

\section{ Mass of the universe and Background Microwave temperature }
\subsection{ Cosmic expansion quantum-cosmon and universe mass }
Now let us calculate the  total energy and the total mass of the universe from the cosmic expansion quanta-cosmons, which are  excitations of the Planckon piled vacuum and their energies come from the lost energies of Planckons during the expansion of the universe. Since the energy ${{e}_{cosmos}}$ of cosmon (cosmic expanson quantum) is the lost energy of Planckon and it contributes to  the universe energy, the lost energy density of Planckon $\Delta\rho_P$ at present time is just the universe energy density $\rho ({{T}_{0}})=\Delta\rho_P$ according to energy conservation. Yet since all Planckons in the vacuum of the expanding universe lose quantum fluctuation energy resulting in negative gravity energy of the vacuum within the universe and the lost energy of Planckons is used to create cosmons which in turn convert into different kinds of universe energy, the total energy of the universe, as the gravity energy is included, namely the negative gravity energy plus the positive different kinds of universe energy is zero. From the Planckon volume  ${{v}_{P}}$ and its lost energy density $\Delta\rho_P=\Delta\rho_{vac}=\rho ({{T}_{0}})$, one obtains the cosmon energy ${{e}_{\cos mos}}$ as follows: ${{e}_{\cos mos}}=\rho ({{T}_{0}})\times {{v}_{P}}$. From $\rho ({{T}_{0}})={{\rho }_{vac}}{{(\frac{{{t}_{P}}}{{{T}_{0}}})}^{2}}$, ${\rho }_{vac}={\rho }_{P}$ , and ${{\varepsilon }_{P}}={{\rho }_{vac}}\times {{v}_{P}}$ , one obtains  ${{e}_{\cos mos}}={{\varepsilon }_{P}}{{(\frac{{{t}_{P}}}{T{}_{0}})}^{2}}\approx {{\varepsilon }_{P}}{{(\frac{{{r}_{P}}}{{{R}_{0}}})}^{2}}\sim {{10}^{-122}}{{\varepsilon }_{P}}\sim {{10}^{-94}}eV$, which is extremely small. Since during the expansion of the universe, each Planckon loses its energy and creates  a  cosmon,  the number ${{N}_{\cos mon}}$  of the cosmons in the universe is just the number ${{N}_{Planckon}}$ of  the Planckons in the universe. According to the Planckon densely piled vacuum model \cite{bib010},
\begin{equation}
{{N}_{\cos mon}}={{N}_{planckon}}=\frac{\frac{4\pi }{3}R_{0}^{3}}{\frac{4\pi }{3}r_{P}^{3}}={{(\frac{{{R}_{0}}}{{{r}_{P}}})}^{3}}\approx {{10}^{183}},  (\frac{{{R}_{0}}}{{{r}_{P}}}=\frac{{{10}^{28}}cm}{{{10}^{-33}}cm}={{10}^{61}})
\end{equation}

The universe  total energy and mass are
\begin{equation}
{{E}_{\cos mos}}={{e}_{\cos mon}}{{N}_{\cos mon}}={{\varepsilon }_{P}}(\frac{{{R}_{0}}}{{{r}_{P}}})={{10}^{80}}GeV, ({{m}_{P}}={{10}^{-5}}g,{{\varepsilon }_{P}}={{10}^{19}}GeV)
\end{equation}
\begin{equation}
{{M}_{\cos mos}}={{m}_{P}}(\frac{{{R}_{0}}}{{{r}_{P}}})={{10}^{-5}}\times {{10}^{61}}g={{10}^{56}}g=5\times {{10}^{22}}{{M}_{\oplus }}, (solar\quad mass \quad {{M}_{\oplus }}=2\times {{10}^{33}}g)
\end{equation}
which agrees very well with the observed value of the universe mass. Thus the universe acquires energy and mass from the Planckon piled vacuum in the form of cosmons which are created simultaneously with the energy loss of Planckons (namely the decrease of vacuum quantum fluctuation energy ) during the expansion of the universe. At a word, the lost part of the vacuum quantum fluctuation energy (the negative hole excitation energy of the Plankon piled vacuum )  manifests itself as the negative gravity potential energy of the vacuum in the universe, while the positive energy of the created cosmons manifests itself as different components of the universe energy.

\subsection{ Temperature of background microwave radiation }

At the decouple time ${{t}_{decouple}}={{10}^{11}}s$,  the photon energy can be calculated as follows:¡¡
\begin{equation}
\rho ({{t}_{decouple}}={{10}^{11}}s)/{{\rho }_{vac}}={{(\frac{{{t}_{P}}}{{{t}_{decouple}}})^{2}}}={{(\frac{{{10}^{-43}}}{{{10}^{11}}})^{2}}}={{10}^{-108}}={{(\frac{{{E}_{decouple}}}{{{\varepsilon }_{P}}})^{4}}}
\end{equation}
\begin{equation}
{{E}_{decouple}}={{10}^{-8}}GeV=10eV\sim{{10}^{5}}{{K}^{0}}.
\end{equation}
The radius of the universe at ${{t}_{decouple}}={{10}^{11}}s$ is
\begin{equation}
R({{t}_{decouple}})={{r}_{P}}(\frac{{{t}_{decouple}}}{{{t}_{P}}})=c{{t}_{decouple}}\sim{{10}^{22}}cm={{R}_{0}}{{10}^{-6}}
\end{equation}
Due to the expansion stretching of the wave length of the photon, the photon energy today is,
\begin{equation}
{{E}_{0}}=\frac{R({{t}_{decouple}})}{R{}_{0}}E({{t}_{decouple}})\sim{{10}^{-1}}{{K}^{0}}
\end{equation}
which is near the observed astronomic value $2.7{{K}^{0}}$ (the error maybe due to the uncertainty of the decouple time or the expansion power laws for
$\rho(t)$ and $R(t)$).

\section{ Different effects of two kinds of vacuum excitaions based on Planckon densely piled vacuum model }
Now we discuss two different kinds of energy losses of Planckons and their different gravity effects. According to the Planckon densely piled vacuum model \cite{bib010}, the black hole gravity is generated by the Casimir effect due to the cutoff of the  black hole horizon surface. Suppose the universe, like a black hole, had a ${{R}_{0}}$-spherical horizon surface which  provides a cutoff for radial wave modes in the Planckon piled vacuum of the universe.  Inside the universe, every Planckon will lose energy from radial modes and the vacuum has less quantum fluctuation energy density than that in flat vacuum. Within the universe, the effective quantum fluctuation temperature of the vacuum  will decrease, which is the microscopic origin of gravity( this is called equilibrium space Casimir effect ). According to \cite{bib010}, the Planckon inside the universe has the energy loss from ${{\varepsilon }_{P}}$,
\begin{equation}
 \Delta {{\varepsilon }_{P}}={{e}_{H}}={{\varepsilon }_{P}}(\frac{{{r}_{P}}}{R{}_{0}})=\frac{\hbar c}{2{{R}_{0}}}
\end{equation}
Just like the Schwarzschild black hole case\cite{bib010}, the energy loss $\Delta {{\varepsilon }_{P}}$ of the Planckon will decrease vacuum quantum fluctuation energy, lead to hole excitations of the Planckon piled vacuum, manifesting as negative gravity potential which will drive particles to the horizon. Like the Schwarzschild black hole, the corresponding excitation quanta of the universe vacuum will be accumulated in horizon in the form of radiation quantum with the energy ${{e}_{H}}$ and the number of the radiation quanta in horizon is just the number ${{N}_{H}}=4{{({{R}_{0}}/{{r}_{P}})}^{2}}$ of the Planckons in the $R_{0}$-spherical horizon Planckon layer. The total mass ${{M}_{m}}$ is the excitation quantum mass ${{e}_{H}}/{{c}^{2}}$ times the number ${{N}_{H}}=4{{({{R}_{0}}/{{r}_{P}})}^{2}}$,
\begin{equation}
{{M}_{m}}=4({{e}_{H}}/{{c}^{2}}){{(\frac{{{R}_{0}}}{{{r}_{P}}})}^{2}}=4{{m}_{P}}(\frac{{{R}_{0}}}{{{r}_{P}}})\sim{\ }{{M}_{\cos mos}}
\end{equation}
which is in the same order of magnitude as the universe mass ${{M}_{\cos mos}}$ .

  On the other hand, the expansion of the universe to the radius ${{R}_{0}}$ leads to the quantum fluctuation energy loss of the Planckons in the universe in quite a different way which can be called as "non-equilibrium time Casimir effect". Since in the expanding universe, the lost energy of each Planckon with volume $v_{P}$ is used to create a cosmon which contributes to the universe with the energy density $\rho ({{T}_{0}})$, the cosmon energy should be
  \begin{equation}
   {{e}_{\cos mon}}=\rho ({{T}_{0}}){{v}_{P}}={{\varepsilon }_{P}}{{(\frac{{{r}_{P}}}{R{}_{0}})}^{2}}=\frac{\hbar c}{2{{r}_{P}}}{{(\frac{{{r}_{P}}}{{{R}_{0}}})}^{2}}=\frac{\hbar c}{2{{R}_{0}}}(\frac{{{r}_{P}}}{{{R}_{0}}})=\frac{\hbar c}{2{{R}_{\kappa }}}, {{R}_{\kappa }}={{R}_{0}}(\frac{{{R}_{0}}}{{{r}_{P}}})
  \end{equation}
They contribute the universe a total mass ${{M}_{\cos mos}}$  as calculated before  ${{M}_{\cos mos}}\sim{{10}^{22}}{{M}_{\oplus }}$.

Now let us investigate the gravity effect due to the creation of cosmons with mean energy ${{e}_{\cos mon}}=\frac{\hbar c}{2{{R}_{\kappa }}}=\frac{\hbar \kappa }{2c}$ and $\kappa =\frac{{{c}^{2}}}{{{R}_{\kappa }}}$. The cosmon is a standing quantum radiation wave with the wave length ${{\lambda }_{\cos mon}}=4\pi {{R}_{\kappa }}$ and spin 1/2 \cite{bib010}. From the period condition of temperature Green's function for fermion, we have ${{e}_{\cos mos}}/{{k}_{B}}{{T}_{\cos mos}}=\pi $,  and the Hawking-Unruh formulae naturally follows $\pi {{k}_{B}}{{T}_{\cos mos}}={{e}_{\cos mon}}=\frac{\hbar \kappa }{2c}=\frac{\hbar c}{2{{R}_{k}}}$ \cite{bib010}. Since the cosmon excitations in the vacuum with positive energy leave holes  with negative energy in the Planckon piled vacuum, the gravity acceleration $\kappa$ and the negative gravity potential $\phi(R)$ resulted from the holes of the vacuum of the universe due to the quantum fluctuation energy loss and the creation of the cosmon, are related as follows: $\kappa =-\frac{d\varphi (R)}{dR}=\frac{{{c}^{2}}}{{{R}_{\kappa }}}$, which  leads to the negative Einstein gravity potential $\varphi (R)=-\frac{{{c}^{2}}}{{{R}_{\kappa }}}R$. Since the above gravity acceleration $\kappa$ is in the radial direction and outwards, the gravity effect induced by the cosmons has the nature of repulsive force. Thus the cosmon as a radiation quantum of fermion-type, maybe  a candidate of the dark energy quantum. It is dual to the radiation quantum for the Schwarzschild black hole \cite{bib010} as follows:

 Cosmon of expanding universe:
 \begin{equation}
  {{e}_{\cos mon}}=\frac{\hbar c}{2{{R}_{\kappa }}},~~~(R{}_{\kappa }={{R}_{0}}({{R}_{0}}/{{r}_{P}})
 \end{equation}

 Radiation quantum of black hole:
 \begin{equation}
{{e}_{H}}=\frac{\hbar c}{2{{r}_{H}}},~~~({{r}_{H}}-black\ hole\ radius)
 \end{equation}

It should be noted that both the universe surface cutoff and the universe expansion lead to quantum fluctuation energy loss of all the planckons in the Planckon piled vacuum of the universe, but the mechanism is quite different. The universe surface cutoff leads to the quantum fluctuation energy loss of every Planckon with the energy decrease of $\Delta {{\varepsilon }_{P}}(space)={{e}_{H}}={{\varepsilon }_{P}}(\frac{{{r}_{P}}}{R{}_{0}})$, which is a violin cord effect
( equilibrium space Casimir effect ); while the universe expansion leads to the quantum fluctuation energy loss of every Planckon with the energy decrease of $\Delta {{\varepsilon }_{P}}(time)={{e}_{\cos mon}}={{\varepsilon }_{P}}{{(\frac{{{r}_{P}}}{R{}_{0}})}^{2}}={{e}_{H}}(\frac{{{r}_{P}}}{{{R}_{0}}})$,
which is ${{r}_{P}}/{{R}_{0}}$ time smaller than $\Delta {\varepsilon_{P}} (space)=e_H$ and shows the effect of outward radiation  with holographic nature (non-equilibrium time Casimir effect). It can be calculated that as standing quantum waves of the cosmons in the universe, their velocity $ v_{cosmon}\sim {\kappa} \times {T_{0}}\sim 10^{-51} {{cm}/{s}}$ and their acceleration $a_{cosmon}={\kappa}=\frac{c^2 r_P}{R^{2}_{0}}\sim 10^{-69}{{cm}/{s^2}}$  are extremely small and negligible, so that the cosmons distribute their  quantum energy ${{e}_{\cos mon}}$ in the universe uniformly and statically. Instead, for the radiation excitations of the universe black hole, their velocity $ v_{H}\sim {\kappa_{H}} \times {T_0} \sim 10^{10}{{cm}/{s}}$ and their acceleration $a_{H}=\kappa_{H}=\frac{c^2}{R_{0}}\sim 10^{-8}{{cm}/{s^2}}$  are sizable and not negligible, so that the radiation quantum moves their energy ${{e}_{H}}$ to the universe surface. Both the expansion cosmons and the radiation quanta are excitations of the vacuum which is densely piled by Planckons.

From the above results and discussions, we come to the very important conclusion that within a spherical region, vacuum quantum fluctuation energy loss necessarily leads to the negative gravity potential and its acceleration is of the nature of a radially outward force: inside the supposed Universe Black Hole, vacuum quantum fluctuation energy loss due to the Casimir effect leads the gravity potential $\phi(R)=-\kappa_H R$ and the gravity acceleration $\kappa_H =\frac{c^2}{R_0} $ is radially outward to the universe horizon, every particle will feel a repulsive force outwards radially; while inside the expanding universe, the vacuum quantum fluctuation energy loss due to the expansion of the universe will definitely induce the gravity potential $\phi(R)=-\kappa R$ and the gravity acceleration $\kappa=\frac{c^2}{R_{\kappa}}=\frac{r_P c^2}{R_0^2} $ is also radially outward, every particle will be pushed radially outwards to the expanding universe horizon. In the two cases. the gravity induced by the vacuum quantum fluctuation energy loss plays a similar role of repulse forces due to the radially outward gravity acceleration.

\section{ Physical explanation of the results }

 The present study tells us that our universe was born from an explosion of a Planckon in the
Planckon densely piled vacuum. The initial condition is thus the Planck era, namely the Planckon space-time scales(${{r}_{P}},{{t}_{P}}$) and its energy density ${{\rho }_{P}}={{\rho }_{vac}}$ are the initial conditions of the Einstein-Friedmann equations. The evolution of the universe is controlled by the Einstein-Fridmann equations with different energy density components. During the evolution, more and more Planckons are involved, lose their quantum fluctuation energy, and create cosmons which change the  lost irregular Planckon energies into their regular energies. The gravity effect due to the creation of cosmon quanta is to produce a repulsive force outward in radial direction, and the cosmon may thus be the candidate of the dark energy quantum.

   The evolution took three steps: (1) from the Planck era to the inflation starting moment, (2)  underwent an inflation with a very short period, (3) from the end of inflation to the present time. Evidently, the evolution of our universe is controlled by the basic principles of general relativity, cosmology, and quantum theory.  Of course,  the information of particle physics is needed to achieve a complete and detailed description.  From our model , the energy density of the dark energy, the inflation phase transition temperature, the background microwave temperature, and the universe total mass can be calculated  in agreement with the observational data.

   An intuitively physical picture can be presented for the time evolution of the universe energy density and the creation of cosmons as follows. In the Planckon densely piled vacuum\cite{bib010}, both the universe and the vacuum constitute a compound system,  each of them are open subsystems. Planckons in the vacuum as radiation sources can radiate and absorb radiation quantum energy. In the stationary universe, the radiation and absorption processes are in equilibrium, there is no net effect can be observed physically in the universe subsystem. As the universe subsystem undergoes expansion, the stationary equilibrium of the two subsystems is broken, every Planckon in the expanding universe is now in non-equilibrium and radiates energy quantum as a cosmon. Therefore a net effect of cosmons in the universe subsystem can be observed physically as outwards radial radiation and the Planckon piled vacuum with less radial quantum fluctuation energy induces a negative gravity potential with the outward strength like a repulsive force.

   Let the pressure of the Planckon at its $r_P$-spherical surface be $p({{r}_{P}})$. As the universe has expanded to ${{R}_{0}}$-sphere,  the radiation pressure of the Planckons at ${{R}_{0}}$ sphere is $p({{R}_{0}})$. Since the  pressure on a spherical surface is proportional to the energy quanta number per unit area and per unit time -the radiation energy flux, according to energy conservation, $p({{R}_{0}})$  is inversely proportional to the spherical surface area ($4\pi R_0^2$). Hence
\begin{equation}
 \frac{p({{R}_{0}})}{p({{r}_{P}})}\sim{{(\frac{{{r}_{P}}}{{{R}_{0}}})}^{2}}, \    p({{R}_{0}})\sim{{(\frac{{{r}_{P}}}{{{R}_{0}}})}^{2}}p({{r}_{P}}),
  \end{equation}

Using the the energy density and pressure relation $p({{r}_{P}})\sim{{\rho }_{vac}}({{r}_{P}})$  and $p({{R}_{0}})\sim\rho ({{R}_{0}})$, finally we have
\begin{equation}
\rho ({{R}_{0}})\sim{{\rho }_{vac}}{{(\frac{{{r}_{P}}}{{{R}_{0}}})}^{2}}\sim{{\rho }_{vac}}{{(\frac{{{t}_{P}}}{{{T}_{0}}})}^{2}}
\end{equation}
This is exactly the results obtained by directly solving the Einstein-Friedmann equations.

  The above picture implies  that: the evolution of the universe energy density,  the creation of cosmons and dark energy,  are the problems of a non-equilibrium and open system,  and the non-equilibrium process  is  more important and essential. Only in its non-equilibrium expansion, the universe can exchange energy with vacuum, the Planckons in the universe can lose energy and change its lost quantum fluctuation energy into cosmon's regular energy, finally the dark energy quanta-cosmons are created. Therefore, the dark energy is not the vacuum quantum fluctuation energy itself, it is the lost energy of Planckons (namely the lost part of vacuum quantum fluctuation energy) under the condition of the universe  non- equilibrium expansion. The cosmon is made of the lost energy of Planckons  and acts as dark energy quantum with the repulsive nature of its gravity effect.

\section{ Summary of the results and discussions }

¡¡The results of this paper are summarized as as follows.\\¡¡
¡¡¡¡1) Based on the Planckon  densely piled model, vacuum quantum fluctuation energy density is that of Planckon  ${{\rho }_{vac}}={{\rho }_{P}}$.\\
    2) By solving the Einstein-Friedmann equations with the Planck era as the initial conditions, in both three steps and one step respectlively,  we have obtained $\rho ({{T}_{0}})/\rho ({{t}_{P}})\sim{\ }{{(\frac{{{t}_{P}}}{{{T}_{0}}})}^{2}}\sim{{10}^{-122}}$. Since the dark energy is $73\%$ of the present universe energy density , so ${{\rho }_{de}}({{T}_{0}})\sim\rho ({{T}_{0}})\sim {{{\rho }_{vac}}{{10}^{-122}}}$.\\
¡¡¡¡3) The relation between dark energy and vacuum quantum fluctuation energy loss is explored and the cosmon with energy ${{e}_{\cos mos}}={{\varepsilon }_{P}}{{(\frac{{{r}_{P}}}{{{R}_{0}}})}^{2}}={{10}^{-122}}{{\varepsilon }_{P}}$ is introduced and proved to be the candidate of the dark energy quantum.\\
 ¡¡4) The calculated inflation phase transition temperature is ${{E}_{c}}\sim{{10}^{15}}GeV$ consistent with the GUT theory. \\
   5) The universe mass calculated is ${{M}_{\cos mos}}\approx {{10}^{22}}{{M}_{\otimes }}$ and the calculated background microwave temperature is $T\sim0.1{{K}^{0}}$ consistent with the observational data. \\
   6) Two different kinds of excitation quanta of the Planckon piled vacuum-the radiation quantum ${{e}_{H}}\sim{{\varepsilon }_{P}}(\frac{{{r}_{P}}}{{{R}_{0}}})$ due to the universe space cutoff Casimir effect and the cosmon ${{e}_{\cos mos}}\sim{{\varepsilon }_{P}}{{(\frac{{{r}_{P}}}{{{R}_{0}}})}^{2}}$ due to the universe expansion  radiation effect are compared and discussed with two different gravity effects. \\
  7) The gravity potential and the gravity acceleration induced from the creation of cosmons are derived and its repulsive force nature has been unveiled.\\
  8) It is shown that our model has two salient features: (i) as the gravity energy is included, the total energy of the unverse, i.e., the negative gravity energy plus the positive universe energy is zero, (ii) the Einstein-Friedmann universe with the Planckon state as its initial condition in its expansion phase and as its smallest limit in its shrinking phase will undergo a cyclic evolution of successive expansion and shrinking.\\
  9) The scaling law of the universe energy density evolution is explored and the radiation holographic nature of the solution is explained as follows:
\begin{equation}
\rho (t-{{t}_{P}})={{\rho }_{vac}}/\{1+(3/2)\frac{1}{{{t}_{P}}}{{[(t-{{t}_{P}})+\int\limits_{{{t}_{P}}}^{t}{w(\tau )d\tau ]}\}^{2}}}
\end{equation}
\begin{align}
&\text{Evolution period:}&  &\text{Planck period}~~ {{t}_{P}}&    &\text{inflation period}~~ {{t}_{\inf }}&  &\text{present time}~~ {{T}_{0}}& \nonumber \\
&\text{Energy density}:&¡¡&{{\rho }_{vac}}\sim\frac{{{\rho }_{vac}}}{{{({{t}_{P}}/{{t}_{P}})}^{2}}}& ¡¡&\rho \sim\frac{{{\rho }_{vac}}}{{{[({{t}_{\inf }}-{{t}_{P}})/{{t}_{P}}]}^{2}}}&¡¡¡¡&\rho \sim\frac{{{\rho }_{vac}}}{{{(c{{T}_{0}}/{{t}_{P}})}^{2}}}&  \nonumber\\
&\text{Time scaling}:& &{{t}_{P}}\to s{{t}_{P}},& &{{t}_{\inf }}\to s{{t}_{\inf }},&          &{{T}_{0}}\to s{{T}_{0}}& \nonumber \\
&\text{Energy density scaling}:&&{{\rho }_{vac}}\to {{\rho }_{vac}}/{{s}^{2}},&&\rho ({{t}_{\inf }})\to \rho ({{t}_{\inf }})/{{s}^{2}},&&\rho ({{T}_{0}})\to \rho ({{T}_{0}})/{{s}^{2}}& \nonumber
\end{align}
The above scaling law is invalid for the evolution of the dark energy density since $\rho (t)={{\rho }_{de}}=const.$

\section{ Relations with other theories and models}
\subsection{ Solutions to three puzzles of the Big Bang model }
 Solutions to three puzzles of the Big Bang model\cite{bib007,bib008} are as follows: \\
(1) The initial singular problem is avoided by the initial conditions of the Planck era. \\
(2) The problem of flatness of the universe space is solved by following calculations:\\
 before inflation, $\dot{R}={{10}^{6}}cm/s$, $|\frac{\rho -{{\rho }_{c}}}{\rho }|\approx \frac{\frac{{{c}^{2}}|k|}{{{R}^{2}}}}{{{H}^{2}}}=\frac{{{c}^{2}}|k|}{{{{\dot{R}}}^{2}}}\sim{{10}^{8}}>>1$, the curvature $k$ term is very large and the space is not flat; after inflation, $\dot{R}={{10}^{49}}cm/s$, $|\frac{\rho -{{\rho }_{c}}}{\rho }|\approx \frac{\frac{{{c}^{2}}|k|}{{{R}^{2}}}}{{{H}^{2}}}=\frac{{{c}^{2}}|k|}{{{{\dot{R}}}^{2}}}\sim{{10}^{-78}}\sim0$, the curvature $k$ term is negligible and the space becomes flat. \\
 (3) The horizon problem is solved by the following calculated results: before inflation the expansion velocity $\dot{R}({{\tau }_{\inf }})\sim{{10}^{6}}cm/s<<c$, the causal connection of the whole universe can be established; after inflation the average radial velocity $\bar{\dot{R}}\approx c $,  the average horizon distance $D$ is approximately equal to the average expansion distance $R$ so that $\frac{D}{R}\approx 1$.
\subsection{ Comparing with Guth theory }
To solve the above difficulties of the Big Bang model, instead of using thermodynamic formulas for energy density, entropy, and temperature within the radiation gas model by Guth\cite{bib007,bib008}, our model is based on the Einstein-Friedmann dynamics with following ingredients: (1) Einstein-Friedmann equations, (2) the initial conditions $({{t}_{P}},{{r}_{P}},{{\rho }_{P}}={{\rho }_{vac}})$  of the Planck era, (3) the equation of state with $w(\rho )=w[\rho (t)]$, and (4) including a short period of inflation.
\subsection{ Comparing with holographic models }
 As discussed in sections IV and V, the dark energy and  its gravity effect are related to  cosmons which are created from the lost quantum fluctuation energies of Planckons during the expansion of the universe.  The dark energy density has the relation $\rho (R)={{\rho }_{vac}}{{(\frac{{{r}_{P}}}{R})}^{2}}$,  just like photons radiated from the source, propagated outwards and left their information at different spherical surfaces. Thus the dark energy with the above property possesses  the nature of holography of radiation, consistent with the holographic models of dark energy\cite{bib011,bib012}.
\subsection{ Different Casimir effects }
 In section IV, we have pointed out that the gravity and temperature of black holes are stemmed from the equilibrium and stationary space Casimir effect \cite{bib010} and the dark energy is from the non-equilibrium and time-dependent casimir effect. The Radiation quantum ${{e}_{H}}$ and  cosmon ${{e}_{\cos mon}}$ are two kinds of basic excitations of the vacuum which is consisted of densely piled Planckons.  Both ${{e}_{H}}$ and ${{e}_{\cos mon}}$ are closely related to the quantum fluctuation energy loss of Planckons in vacuum under different conditions: the space boundary cutoff of the supposed universe black hole causes the space Casimir effect and induces the radiation quantum ${{e}_{H}}$,  while the expansion of the universe causes the time Casimir effect and induces the cosmon ${{e}_{\cos mon}}$. Since these two kinds of basic effects are related to fundamental and important physics, the issue of different Casimir effects deserves further study.

\section*{ Acknowledgements }
This work was supported in part by the National Natural Science Foundation of China under the grant No.\ 10974137 and 10775100,  the Doctoral Education Fund of the Education Ministry of China, and by the Research Fund of the Nuclear Theory Center of HIRFL of China.


\begin{thebibliography}{20}
\bibitem{bib001}
A.~G.~Riess {\it et al.}  [Supernova Search Team Collaboration],
Astron.\ J.\  {\bf 116} (1998) 1009
\bibitem{bib002}
S.~Perlmutter,{\it et et al},Nature, 391 51(1998); Astrophys. Journ. 517, 565(1999); astro-ph/9712212.
\bibitem{bib003}
S.~Dodelson,[Modern Cosmology]~(Academic Press, 2003)
\bibitem{bib004}
Q.L.Lu£¬\ Z.L. Chou£¬\ H. Y. Guo£¬¡¶Acta Physica(China)¡·£¬23(1974) 225;
H. Y. Guo£¬¡¶Science Bulletin¡·£¬22(1977) 481
\bibitem{bib005}
H. Y. Guo,\ C. G. Huang,\ Z. Xu, and B. Zhou, Mod.Phys.Lett. A19, 1701(2004).
\bibitem{bib006}
 H. Y. Guo,\ C. G. Huang,\ Z. Xu, and B. Zhou, Phys.Lett. A331,1(2004).
\bibitem{bib007}
A. Guth, Phys. Rev. D23, 347(1981)
\bibitem{bib008}
A. Lindle, Phys. Lett. B108, 389(1982)
\bibitem{bib009}
S. Weinberg,  Rev. Mod. Phys. 61, 1(1989)
\bibitem{bib010}
Shun-Jin Wang,~Microscopic quantum structure of black hole and vacuum versus  quantum statistical origin of gravity, arXiv:1212.5862[gr-qu]v4, Oct. 28. 2014£®
\bibitem{bib011}
A. Cohen, D. K,aplan, A. Nelson,   Phys. Rev. Lett. 82(1999) 4971.
\bibitem{bib012}
Miao Li,  Phys.Lett.B603,  (2004) 1-5
\end{thebibliography}
\end{document}